# The nonextensive parameter and Tsallis distribution for self-gravitating systems

Du Jiulin

*Department of Physics, School of Science, Tianjin University, Tianjin 300072, China*

E-mail: jldu@tju.edu.cn

**Abstract**

The properties of the nonextensive parameter $q$ and the Tsallis distribution for self-gravitating systems are studied. A mathematical expression of $q$ is deduced based on the generalized Boltzmann equation, the $q$-$H$ theorem and the generalized Maxwellian $q$-velocity distribution in the framework of Tsallis statistics. We obtain a clear understanding of the physics of $q \neq 1$ with regard to the temperature gradient and the gravitational potential of the self-gravitating systems. It is suggested that the Tsallis statistics could be statistics suitable for describing the nonequilibrium systems with inhomogeneous temperature and long-range interactions.

In the framework of Boltzmann-Gibbs (B-G) statistical mechanics, the structure and stability of self-gravitating systems at statistical equilibrium are usually analyzed in terms of the maximization of a thermodynamic potential. This thermodynamic approach leads us to the isothermal configurations that have been studied for a long time in the context of stellar structure and galactic structure. The equilibrium properties of self-gravitating systems are described by the equation of state of an idea gas taken in the form of $P = nkT$ and the Maxwell-Boltzmann (M-B) distribution [1,2] given by

$$f(\mathbf{r},\mathbf{v}) = \left(\frac{m}{2\pi kT}\right)^{\frac{3}{2}} n(\mathbf{r}) \exp\left(-\frac{m\mathbf{v}^2}{2kT}\right), \qquad (1)$$

where the temperature $T$ is a constant and the density distribution is

$$n(\mathbf{r}) = n_0 \exp[-\frac{m}{kT}(\varphi - \varphi_0)], \qquad (2)$$

where $n_0$ and $\varphi_0$ are the number density of particles and the gravitational potential at $r = 0$, respectively. The gravitational potential $\varphi$ satisfies the Poisson equation,

$$\nabla^2 \varphi = 4\pi G m n. \qquad (3)$$

However, it is well known that self-gravitating systems are usually at a hydrostatic equilibrium and is not at a thermal equilibrium. Almost all the self-gravitating systems at stable states are always non-isothermal. The isothermal configurations only correspond to the meta-stable states (locally convective mixing), not true equilibrium states. Therefore, the M-B distribution (1) and the density distribution (2) cannot provide an exact description of the properties of self-gravitating systems because they ignore the spatial inhomogeneity of temperature. Additionally, self-gravitating systems are open systems at nonequilibrium states. The conventional statistical mechanics of self-gravitating systems has exhibited peculiar features such as infinite mass problem, gravothermal catastrophe, negative specific heats and inequivalence of statistical ensembles [2,3], which are greatly different from the usual thermodynamic systems and so quite difficult to understand in the B-G statistical framework.

It has been considered that the systems with long-range interactions are nonextensive and the conventional B-G statistical mechanics may be not appropriate to the description of the features of the systems. A nonextensive generalization of B-G statistical mechanics known as "Tsallis statistics" has attracted great attention in recent years [4]. This generalization was made by constructing a new form of entropy $S_q$ to be related to a parameter $q$ different from unity [5] by

$$S_q = \frac{k}{1-q}(\sum_i p_i^q - 1), \tag{4}$$

where $k$ is the Boltzmann constant, $p_i$ is the probability that the system under consideration is in its $i$th configuration, and $q$ is the nonextensive parameter whose deviation from unity is considered as describing the degree of nonextensivity of the system. The Boltzmann entropy $S_B$ is recovered from $S_q$ only if $q=1$. In this way, Tsallis statistics gives a power law distribution for all $q \neq 1$, while the B-G exponential distribution is obtained only when $q=1$. This new theory has provided a convenient frame for the thermo-statistical analyses of many astrophysical systems and processes, such as stellar polytropes [6-8], the gravothermal catastrophe [7,8], the negative specific heat [9,10], the velocity distribution of galaxy clusters [11], the instability of self-gravitating systems [12,13], and the solar neutrino problem [14] etc. It has been reported [11] that the observed velocity distribution of galaxy clusters significantly deviates from the M-B distribution and can be fitted well by the Tsallis distribution with $q = 0.23^{+0.07}_{-0.05}$. By performing a set of numerical simulation of the long-term stellar dynamical evolution, it has been found [15] that the quasi-equilibrium sequence with different $q$ values arising from the Tsallis entropy plays an important role in characterizing the transient state away from the B-G equilibrium state in the nonequilibrium dynamical evolution of stellar self-gravitating systems. These work made it clear that Tsallis statistics may be the theory appropriate to the description of astrophysical systems with the long-range interactions of gravitation.

However, the true nature of the nonextensive parameter $q$ in Tsallis statistics has not yet to be well understood [16]. It is unclear under what circumstances, e.g. which class of nonextensive systems and under what physics situation, should Tsallis statistics be used for the statistical description. The understanding of the physical meaning of $q$ has become crucially important in Tsallis statistics and its applications to the fields of astrophysics. In this Letter, we study the properties of the nonextensive parameter $q$ and the Tsallis distribution of the self-gravitating systems and deduce a mathematical expression of $q$ based on the theory of the generalized Boltzmann equation and the $q$-$H$ theorem in the nonextensive context of Tsallis statistics. We will discuss the physics of $q$ and try to answer the question of "what the nonextensive parameter $q \neq 1$ stands for in the self-gravitating systems?" In addition, we also obtain the generalized form of the M-B distribution for the self-gravitating systems.

We consider a system of $N$ particles in the $q$-generalized kinetic theory, interacting via the Newtonian gravitation $\mathbf{F} = -\nabla \varphi$. The mass of each particle is $m$. We let $f_q(\mathbf{r}, \mathbf{v}, t)$ be the distribution function of particles in the self-gravitating system, and then $f_q(\mathbf{r}, \mathbf{v}, t) d^3\mathbf{r} d^3\mathbf{v}$ is the particle numbers at time $t$ and in the volume element $d^3\mathbf{r} d^3\mathbf{v}$ around the position $\mathbf{r}$ and the velocity $\mathbf{v}$. The dynamical behavior of the self-gravitating system is governed by the generalized Boltzmann equation [17],

$$\frac{\partial f_q}{\partial t} + \mathbf{v} \cdot \frac{\partial f_q}{\partial \mathbf{r}} - \nabla \varphi \cdot \frac{\partial f_q}{\partial \mathbf{v}} = C_q(f_q), \tag{5}$$

where the gravitational potential $\varphi$ is determined by the Poisson equation (3) and $C_q$ is called the $q$-collision term. It has been verified [17] that solutions of the generalized Boltzmann equation (5) satisfy the generalized $q$-$H$ theorem only if $q > 0$ and evolve irreversibly towards the Tsallis' equilibrium distribution (the generalized Maxwellian $q$-velocity distribution [9,17,18,19]). The extension of the Tsallis distribution to include nonuniform systems with interparticle interactions results in

$$f_q(\mathbf{r}, \mathbf{v}) = n B_q \left(\frac{m}{2\pi kT}\right)^{\frac{3}{2}} \left[1 - (1-q)\frac{m(\mathbf{v} - \mathbf{v}_0)^2}{2kT}\right]^{\frac{1}{1-q}}, \tag{6}$$

where $\mathbf{v}_0$ is the macroscopic entirety-moving velocity of the system, $n$ is the number density of particles, $T$ is the temperature ($n$ and $T$ are now functions of the space coordinate $r$), and $B_q$ is a $q$-dependent normalized constant given by

$$B_q = \frac{1}{4}(1-q)^{\frac{1}{2}}(3-q)(5-3q)\Gamma\left(\frac{1}{2}+\frac{1}{1-q}\right)\Big/\Gamma\left(\frac{1}{1-q}\right) \quad \text{for} \quad q \leq 1, \tag{7}$$

$$B_q = (q-1)^{\frac{3}{2}}\Gamma\left(\frac{1}{q-1}\right)\Big/\Gamma\left(-\frac{3}{2}+\frac{1}{q-1}\right) \quad \text{for} \quad q \geq 1. \tag{8}$$

Usually, the self-gravitating system at a stable state is believed to satisfy the condition of hydrostatic equilibrium and it is considered as being at a relatively static state. In this case, the macroscopic entirety-moving velocity is zero, $\mathbf{v}_0 = 0$, and Eq.(6) becomes

$$f_q(\mathbf{r},\mathbf{v}) = nB_q\left(\frac{m}{2\pi kT}\right)^{\frac{3}{2}}\left[1-(1-q)\frac{m\mathbf{v}^2}{2kT}\right]^{\frac{1}{1-q}}. \tag{9}$$

In this distribution function, there is a thermal cutoff on the maximum value allowed for the velocity of a particle for $q<1$, $v_{\max} = \sqrt{2kT/m(1-q)}$, while there is no thermal cutoff for $q>1$. This generalized Maxwellian $q$-velocity distribution has been used to investigate the negative heat capacity [9], Jeans criterion of the self-gravitating system [12,13], the non-Maxwellian velocity distribution for the plasma system [19], the nonextensive transport property [20], the nonextensive distribution in a conservative force field [21] and so on. For convenience, let $Q = 1-q$, we write Eq.(9) as

$$f_Q(\mathbf{r},\mathbf{v}) = nB_Q\left(\frac{m}{2\pi kT}\right)^{\frac{3}{2}}\left[1-Q\frac{m\mathbf{v}^2}{2kT}\right]^{\frac{1}{Q}}. \tag{10}$$

We now consider the generalized Boltzmann equation. When the $q$-$H$ theorem is satisfied, we have $C_q = 0$ and $\frac{\partial f_q}{\partial t} = 0$ and so Eq.(5) is reduced to

$$\mathbf{v}\cdot\nabla f_q - \nabla\varphi\cdot\nabla_v f_q = 0, \tag{11}$$

where we have used $\nabla = \partial/\partial\mathbf{r}$ and $\nabla_v = \partial/\partial\mathbf{v}$. The properties of the Tsallis' equilibrium for the self-gravitating system can be analyzed by Eq.(11). We use $Q=1$-$q$ again and write Eq.(11) as

$$\mathbf{v}\cdot\nabla\ln f_Q - \nabla\varphi\cdot\nabla_v \ln f_Q = 0. \tag{12}$$

In the following development, the parameter $Q$ is considered as being a function of $r$ in order to obtain a more detailed description of nonextensivity within the system. From Eq.(10) we obtain

$$\ln f_Q = \ln\left[nB_Q\left(\frac{m}{2\pi kT}\right)^{\frac{3}{2}}\right] - \sum_{i=1}^{\infty}\frac{1}{i}Q^{i-1}\left[\frac{mv^2}{2kT}\right]^i. \tag{13}$$

This expansion of power series is allowed only for the condition $-1 \leq Qmv^2/2kT < 1$ to be satisfied, which is equivalent to $v < \sqrt{2kT/mQ}$ for $Q>0$ and $v \leq \sqrt{-2kT/mQ}$ for $Q<0$. This condition is actually the thermal cutoff on the maximum value allowed for the velocity of particles for $q<1$, $v < v_{max} = \sqrt{2kT/m(1-q)}$, and for $q>1$, $v \leq v_{max} = \sqrt{2kT/m(q-1)}$. Substitute Eq.(13) into Eq.(12), we have

$$\mathbf{v}\cdot\left\{\nabla\ln\left[nB_Q\left(\frac{m}{2\pi kT}\right)^{\frac{3}{2}}\right] + \sum_{i=1}^{\infty}\left(Q^{i-1}\frac{\nabla T}{T} + \left(\frac{1}{i}-1\right)Q^{i-2}\nabla Q\right)\left(\frac{mv^2}{2kT}\right)^i\right\}$$

$$+ \nabla\varphi\cdot\frac{m\mathbf{v}}{kT}\sum_{i=0}^{\infty}\left(\frac{Qmv^2}{2kT}\right)^i = 0. \tag{14}$$

Because $\mathbf{r}$ and $\mathbf{v}$ are independent variables and Eq.(14) is identically null for any arbitrary $\mathbf{v}$, the coefficients of the powers of $\mathbf{v}$ in Eq.(14) must be zero. The properties of Tsallis' equilibrium for the system can be determined by these equations of the coefficients [22]. First, we consider the coefficient equation for the first-order terms of $\mathbf{v}$ in Eq.(14) and we find

$$\nabla\ln\left[nB_Q\left(\frac{m}{2\pi kT}\right)^{\frac{3}{2}}\right] + \frac{m\nabla\varphi}{kT} = 0. \tag{15}$$

After the integral for Eq.(15) is completed, we obtain the temperature-dependent density distribution,

$$n(\mathbf{r}) = n_0\frac{B_{Q(r=0)}}{B_{Q(\mathbf{r})}}\left(\frac{T(\mathbf{r})}{T_0}\right)^{\frac{3}{2}}\exp\left[-\frac{m}{k}\left(\int\frac{\nabla\varphi(\mathbf{r})}{T(\mathbf{r})}\cdot d\mathbf{r} - \frac{\varphi_0}{T_0}\right)\right], \tag{16}$$

where $n_0, T_0, \varphi_0$ are the integral constants and they denote the density, the temperature and the gravitational potential at $r = 0$, respectively. Different from the distribution Eq.(2), Eq.(16) depends on the temperature $T(r)$ and the $q$-dependent normalized constant $B_Q$.

Second, we consider the coefficient equation for the third-order terms of **v** in Eq.(14) and we find the relation

$$k\nabla T + Qm\nabla \varphi = 0. \tag{17}$$

It is shown clearly in this relation that the nonextensive parameter is $Q \neq 0$ if and only if the temperature gradient is $\nabla T \neq 0$, which gives a clear physics of $q \neq 1$ with regard to the nature of non-isothermal configurations of the self-gravitating system. Furthermore, we can write Eq.(17) as

$$Q = -\frac{k\nabla T \cdot d\mathbf{r}}{m\nabla \varphi \cdot d\mathbf{r}}. \tag{18}$$

This gives an explicit mathematical expression of the nonextensive parameter $q$.

Third, we consider the coefficient equation for the fifth-order terms of **v** in Eq.(14). It is given by

$$\left(\frac{m}{2kT}\right)^2 \left(Q\nabla T - \frac{1}{2}\nabla Q + Q^2 \frac{m}{k}\nabla \varphi\right) = 0. \tag{19}$$

We get from Eq.(19)

$$\nabla Q = 0. \tag{20}$$

It can be verified that all the coefficient equations of powers of **v** will lead to Eq.(18) and Eq.(20). Eq.(18) or Eq.(17) is the only mathematical expression that has been reported so far for the nonextensive parameter $q$, and it presents $q \neq 1$ a clear physics meaning with regard to the temperature gradient and the gravitational potential. If the temperature gradient is $\nabla T = 0$, then we have $Q = 0$ and $q=1$, which corresponds to the case of B-G statistics, while if the temperature gradient is $\nabla T \neq 0$, then we have $Q \neq 0$ and $q \neq 1$, which corresponds to the case of Tsallis statistics. Therefore, the nonextensive parameter $q \neq 1$ is shown to be responsible for the spatial inhomogeneity of temperature as well as the gravitational long-range interactions and it is rationally related to the non-isothermal nature of the self-gravitating system at the nonequilibrium stationary state.

Due to Eq.(20), we have $B_{Q(r)} = B_{Q(r=0)}$. Substitute Eq.(16) into Eq.(10), we can write the generalized M-B $q$-distribution (or the Tsallis distribution) for the self-gravitating system as

$$f_Q(\mathbf{r},\mathbf{v}) = n_0 B_Q \left(\frac{m}{2\pi k T_0}\right)^{\frac{3}{2}} \exp\left[-\frac{m}{k}\left(\int \frac{\nabla \varphi(\mathbf{r})}{T(\mathbf{r})}\cdot d\mathbf{r} - \frac{\varphi_0}{T_0}\right)\right]\left[1 - Q\frac{m\mathbf{v}^2}{2kT(\mathbf{r})}\right]^{\frac{1}{Q}}, \quad (21)$$

where $T_0$ is naturally equivalent to $T$ given in Eq.(1) because, in Eq.(1), the temperature is a constant. The important difference between Eq.(21) and Eq.(1) is that Eq.(21) contains the parameter $Q$ and so includes the contributions of temperature gradient $\nabla T$. Due to this characteristic, the generalized M-B $q$-distribution (21) could describe the nature of non-isothermal configurations of the self-gravitating systems at the nonequilibrium stationary states. It is clear that the M-B distribution Eq.(1) is recovered perfectly from Eq.(21) if we take $\nabla T = 0$. In this way, it is very possible to find the experimental evidence for a value of $q$ different from unity and then to test the theory of nonextensive statistical mechanics.

We make a comparison between Eq.(18) and the theorem proved by Almeida recently. Almeida's theorem [23] states that the canonical distribution function of a system is Tsallis distribution if and only if the equation

$$\frac{d}{dE}\left(\frac{1}{\beta}\right) = q - 1 \quad (22)$$

with $\beta = 1/kT$ is satisfied, where $T$ is the equilibrium temperature and $E$ is the energy of the environment (or the "heat bath"). Some of authors have given $q \neq 1$ an interpretation about finite heat capacity of the "heat bath", leading Tsallis statistics to a small "heat bath" statistics [24]. We might extend Eq.(22) to nonuniform systems. If the environment around the system is space inhomogeneous, rather than the uniform "heat bath" as usual, then the energy $E$ and the temperature $T$ are considered as being a function of $\mathbf{r}$. From Eq.(22) we directly obtain

$$1 - q = -k\frac{\nabla T \cdot d\mathbf{r}}{\nabla E \cdot d\mathbf{r}}. \quad (23)$$

This relation is obviously consistent with Eq.(18). It means that Tsallis statistics might be a theory suitable for describing the properties of the systems with inhomogeneous temperature and with long-range potential. As is well known, astrophysical systems have the environment of the gravitational long-range interactions. According to the Virial theorem, half of the gravitational energy is transformed into heat when the system contracts by the self-gravitating force. The spatial inhomogeneity of astrophysical environment leads to the spatial inhomogeneity of the temperature and the density in the self-gravitating system. In fact, in a nonequilibrium open system, the system and its environment exchange energy and matter between each other constantly.

Recent investigations on nonextensive statistical mechanics for self-gravitating systems lead to the power-law distribution referred to as the stellar polytrope, which relates $q$ to the polytrope index n as n = 3/2 + 1/($q$-1) depending on the standard statistical averages [6,7] or n =1/2+ 1/(1-$q$) depending on the normalized $q$-averages [8]. It is worth to notice that the limit $q \to 1$ (or n$\to \infty$) is corresponding to the isothermal distribution of the system. In addition, Boghosian et al [25] relate $q$-1 to the number of dimensions in Lattice Boltzmann models, and Beck [26] relates it to the number of degrees of freedom contributing to a fluctuating spatio-temporal temperature field.

In conclusion, a mathematical expression of the nonextensive parameter $q$ have been deduced based on the generalized Boltzmann equation, the $q$-$H$ theorem and the generalized Maxwellian $q$-velocity distribution in the framework of Tsallis statistics. We have obtained a clear understanding of the physics of $q \neq 1$ related to the temperature gradient and the gravitational potential of the self-gravitating system. We also propose that Tsallis statistics could be the statistics suitable for describing the nonequilibrium properties of the systems with inhomogeneous temperature and long-range interaction potential.

**Acknowledgments**

I would like to thank Dr. W-S Dai for useful discussions and Dr. X-Sh Liu for his help. This work is supported by the project of "985" Program of TJU of China.